\documentclass[twocolumn,english,amsart,showpacs,preprintnumbers,amsmath,amssymb,floatfix]{revtex4-1}
\usepackage{tikz,xcolor}
\usepackage{amsmath}
\usepackage[colorlinks = true,
linkcolor = blue,
urlcolor  = blue,
citecolor = blue,
anchorcolor = blue]{hyperref}

\definecolor{lime}{HTML}{A6CE39}
\DeclareRobustCommand{\orcidicon}{%
	\begin{tikzpicture}
		\draw[lime, fill=lime] (0,0)
		circle [radius=0.16]
		node[white] {{\fontfamily{qag}\selectfont \tiny ID}};
		\draw[white, fill=white] (-0.0625,0.095)
		circle [radius=0.007];
	\end{tikzpicture}
	\hspace{-2mm}
}

\foreach \x in {A, ..., Z}{%
	\expandafter\xdef\csname orcid\x\endcsname{\noexpand\href{https://orcid.org/\csname orcidauthor\x\endcsname}{\noexpand\orcidicon}}
}

\usepackage[T1]{fontenc}
\usepackage[latin9]{inputenc}
\usepackage{color}
\usepackage{array}
\usepackage{amstext}
\usepackage{graphicx}
\usepackage{esint}
\usepackage{rotating}
\usepackage{appendix}
\usepackage{float}

\usepackage[font=small,labelfont=bf]{caption}
\usepackage{xcolor}
\usepackage{ulem}

\makeatletter

\@ifundefined{textcolor}{}
{%
	\definecolor{BLACK}{gray}{0}
	\definecolor{WHITE}{gray}{1}
	\definecolor{RED}{rgb}{1,0,0}
	\definecolor{GREEN}{rgb}{0,1,0}
	\definecolor{BLUE}{rgb}{0,0,1}
	\definecolor{CYAN}{cmyk}{1,0,0,0}
	\definecolor{MAGENTA}{cmyk}{0,1,0,0}
	\definecolor{YELLOW}{cmyk}{0,0,1,0}
}

\@ifundefined{definecolor}
{\usepackage{color}}{}
\@ifundefined{definecolor} 
{\usepackage{color}}{}
\makeatother
\usepackage{babel}

\allowdisplaybreaks

\begin{document}
	
	%%%%%%%%%%%%%%%%%%%%%%%%%%%%%%

	\title{ Examining the structure functions of nucleons  by introducing new Ansatz and their consequences on parton distribution functions in higher approximation
	 }
	
	\author{Hossein Vaziri\orcidA{}}
    \email{Hossein.Vaziri@shahroodut.ac.ir
    }
	
	\author {Mohammad Reza Shojaei\orcidB{}}
	\email{Shojaei.ph@gmail.com}

\affiliation {
	Department of Physics, Shahrood University of Technology, P. O. Box 36155-316, Shahrood, Iran }

	\date{\today}

	%
	%%%%%%%%%%%%%%%%%%%%%%%%%%%%%%%%%%%%%%%%%%%%%  Abstract    %%%%%%%%%%%%%%%%%%%%%%%%%%%%%%%%%%%%%%%%%%%%%%%%%%%%%%%%%%
	%

	%
	\begin{abstract}\label{abstract}
	By introducing ansatz and using a PDF at the $N^3LO$ approximation in this paper, we have studied the structure of the nucleons in generalized parton distributions (GPDs). For this purpose, the form factors and the electric radius of nucleons, as well as the quark's form factors, are calculated in this paper. These calculations give important new information about the nucleons and the distribution of partons within them. The results are analyzed, and to validate them, the abstract emphasizes the importance of comparing the results with existing research and experimental data. By varying the free appropriate parameters of the new ansatz, the study seeks to optimize the model, ensure its suitability for accurately describing the nucleon's internal structure, and provide valuable insights about the distribution of partons.
	\end{abstract}
	%

	%\pacs{12.38.-t, 24.85.+p, 13.15.+g, 13.60.Hb}
	
	\maketitle

	%
	%%%%%%%%%%%%%%%%%%%%%%%%%%%%%%%%%%%%%%%%%%    Introduction    %%%%%%%%%%%%%%%%%%%%%%%%%%%%%%%%%%%%%%%%%%%%%%%%%

	\section{INTRODUCTON}\label{sec:sec1}
In the study of particle physics, collisions and scattering are the most important processes to investigate the particle structure. Using scattering, we study the internal structure of atoms, nuclei, and nucleons. We have been seeking to obtain a connection between the theoretical and experimental worlds, and since the first step in establishing this relationship is to obtain the cross-section, that is required in order to calculate the scattering cross-section. Using deep inelastic scattering (DIS) data of an electron off a proton
(and more generally a nucleon), we can determine their structure functions. The structure function is a mathematical function that depends on the variable $x$ B-Jorken, which indicates the distribution of the percentage of transferred momentum between the partons~\cite{Halzen,Radyushkin:2011dh,Rosenbluth:1950yq,Arnold:1980zj,JeffersonLabHallA:2011yyi,Gramolin:2021gln,Guidal:2004nd,Mondal:2015uha,RezaShojaei:2016oox,Burkardt:2002hr,Miller:1990iz,Guidal:2013rya,Chakrabarti:2014gya,Adhikari:2016idg}. Then, parton distribution functions (PDFs) are calculated based on them. PDFs are functions that represent the two-dimensional distribution of parton positions within nucleons and are functions of $x$, which is the percentage of the nucleus' momentum that the quarks carry after scattering, and it varies between 0 and 1, and the momentum transfer between the last and first nucleons is denoted by $t=Q^2$. GPDs demonstrate the distribution of quarks and gluons in momentum space inside the hadrons and are a theoretical framework in quantum chromodynamics (QCD) used to describe the internal structure of hadrons. They provide information about the distribution of quarks and gluons in both position and momentum space within these particles in three dimensions. They offer valuable insights into parton dynamics and contribute to our understanding of strong interactions and the structure of nucleons, and in addition to $x$ and $t$, it depends on $\xi$. The momentum transmitted to the nucleus has both longitudinal and transverse components, with the fraction of longitudinal momentum transfer indicated by $\xi$. Exploring the generalized parton distributions with regard to the two variables $x$ and $\xi$ can provide us with a wealth of insights into the features of the nucleon.~\cite{Vaziri:2023xee, Bhattacharya:2019cme, Muller:1994ses}.
Nucleon structure can be calculated for valuable information, such as their radius and electromagnetic form factors and  the transversely unpolarized densities for the nucleons, using GPDs. So far, several studies have been done in the field of GPD at lower approximations~\cite{Jimenez-Delgado:2008orh,Bhattacharya:2019cme, Muller:1994ses, Selyugin:2009ic, Hou:2019efy, Martin:2002dr, Vaziri:2023xee, HajiHosseiniMojeni:2022tzn,Nikkhoo:2015jzi}.
Ansatz that shows how $x$ and $t$ depend on generalized parton distributions (GPDs) is introduced in this section. We select a parton distribution function in the $N^3LO$ approximation, demonstrating favorable agreement with experimental data for form factors and radii calculations's of the nucleons.

The content of the present paper is as follows: In Sec.\ref{sec:sec2}, we introduce the formalism of the GPD and an Ansatz. The magnetic and electric form factors of the nucleons included in the present study and related remarks are presented in Sec.\ref{sec:sec4}. Sec. \ref{sec:sec5}. is devoted to presenting the electric radii of the nucleons based on VS24 Ansatz.  We summarize our results and conclusions in Sec.\ref{sec:conclusion}.

      \section{THE FORMALISM OF THE GPD AND INTRODUCTION OF AN ANSATZ}\label{sec:sec2}
The ability to resolve individual quarks and gluons makes deep inelastic scattering processes a critical instrument for hadron structure studies. The data obtained through these processes encapsulates the longitudinal momentum and polarization distribution transported by quarks, antiquarks, and gluons within a high-speed hadron. This information has played a pivotal role in shaping our understanding of the physical aspects of hadron structure. Nonetheless, these measurements omit crucial details, especially regarding the spatial distribution of partons in the transverse plane relative to the hadron's direction of motion and the significant role of their orbital angular momentum in contributing to the overall nucleon's spin. In recent times, it's now evident that specific exclusive scattering processes offer such information, encoded within the  generalized parton distributions (GPDs). 
Generalized parton distributions (GPDs) are considered to be one of the crucial methods for analyzing the structure of nucleons~\cite{Diehl:2003ny,Shojaei:2015oia,SattaryNikkhoo:2018odd}. GPDs are a set of non-perturbative QCD functions that provide a more comprehensive description of hadron structure compared to traditional parton distribution functions (PDFs), which describe the probability of finding quarks and gluons with a specific momentum fraction within a hadron. GPDs contain specific details on the internal structure of hadrons compared to PDFs. Transition matrix elements among states with differing momenta are characterized as GPDs~\cite{SattaryNikkhoo:2018odd,Diehl:2003ny}. The Dirac and Pauli form factors $F_1(t)$ and $F_2(t)$ parameterize the proton matrix element of this momentum.

GPDs provide a more comprehensive understanding of the internal structure of hadrons compared to PDFs.
	\begin{equation}
		F_{1}(t)=\sum_{q}e_{q}\int_{-1}^{1}dx{H}_q(x,t,\xi),
		\label{eq:F1}
	\end{equation}

\begin{equation}
F_{2}(t)=\sum_{q}e_{q}\int_{-1}^{1}dx{E}_q(x,t,\xi).
	\label{eq:F2}
\end{equation}

 When the momentum is transverse and located in the space-like region, the value of $\xi$ is equal to zero. Researchers have revised elastic form factors to reduce the integration region within the $0 < x < 1$ range, resulting in alternative expressions~\cite{HajiHosseiniMojeni:2022okc,Mojeni:2020rev}:
	\begin{equation}
		F_{1}(t)=\sum_{q}e_{q}\int_{0}^{1}dx\mathcal{H}^{q}(x,t),
		\label{eq:F3}
	\end{equation}

	\begin{equation}
		F_{2}(t)=\sum_{q}e_{q}\int_{0}^{1}dx {\varepsilon }^{q}(x,t).
		\label{eq:F4}
\end{equation}

Experimentally, the proton helicity-flip form factor exhibits a faster reduction at large $t$ than does the helicity-conserving form factor. In various models, distinctions arise between the functions $\mathcal{H}(x)$ and $\varepsilon(x)$. Specifically, in the limit as $x$ approaches 1, the function $\varepsilon(x)$ should exhibit a more rapid reduction with $t$ compared to $\mathcal{H}(x)$. This behavior is achieved by incorporating additional powers of $(1-x)$ in the expression for $\varepsilon(x)$~\cite{Guidal:2004nd,Selyugin:2009ic}. Hence, it can be inferred that:
\begin{eqnarray}
	\varepsilon _{u}(x) &=&\frac{\kappa _{u}}{N_{u}}(1-x)^{\eta _{u}}u_{v}(x), \nonumber\\
	\varepsilon _{d}(x) &=&\frac{\kappa _{d}}{N_{d}} (1-x)^{\eta _{d}}d_{v}(x).\label{eq:Eud1}
\end{eqnarray}
\begin{equation}
	\kappa _{q}=\int_{0}^{1}dx\varepsilon _{q}(x).
\end{equation}
The constraints on $\kappa_{q}$ require the proton value to be $F_2^P(0)=\kappa_p = 1.793$ and the neutron value to be $F_2^n(0)=\kappa_n=-1.913$. 
\begin{eqnarray}
	\kappa_u=\kappa_n+2\kappa_p,\nonumber\\
	\kappa_d=2\kappa_n+\kappa_p.
\end{eqnarray}
These considerations lead to the determination of specific parameters. For instance, the values $\kappa_u = 1.673$ and $\kappa_d = -2.033$ can be obtained. Moreover, the normalization integral for the mathematical constant $\int_{0}^{1}\mathcal{H}_{q}(x,0)$ takes on particular values for the nucleons: $F^P_1(0) = 1$ for the proton and $F_1^n(0) = 0$ for the neutron. The normalization integrals for the $\mathcal{H}^{u}(x)=u_v(x)$ and $\mathcal{H}^{d}(x)=d_v(x)$ distributions, which correspond to the valence quark numbers for the $u$ and $d$ quarks, respectively, are 2 and 1. This information allows for the determination of the calculated normalization factors $N_u$ and $N_d$~\cite{Guidal:2004nd}:
\begin{eqnarray}
	N_{u} &=&\int_{0}^{1}dx(1-x)^{\eta _{u}}u_{v}(x), \\
	N_{d} &=&\int_{0}^{1}dx(1-x)^{\eta _{d}}d_{v}(x).  \nonumber
	\label{eq:F5}
\end{eqnarray}

We have fitted the nucleon form factor data to find the values of $\eta_u$ and $\eta_d$ to satisfy the conditions stated in Eq. (\ref{eq:Eud1}). In this paper, several ansatzes with a $t$ dependency will be discussed: the extended ER ansatz~\cite{Guidal:2004nd}, modified Gaussian (MG) ansatz~\cite{Selyugin:2009ic}, HS22 ansatz~\cite{HajiHosseiniMojeni:2022tzn} and M-HS22~\cite{Vaziri:2023xee}.

 In the following sections, we will explore the implications of these ansatzes and compare their predictions with experimental results, aiming to enhance our understanding of GPDs and their connection to the nucleon's form factors. 
 
   The ER ansatz, is~\cite{Guidal:2004nd}: 
 
 \begin{equation}
 	\mathcal{H}^{q}(x,t)=q_{v}(x)x^{-\alpha ^{\prime }(1-x)t},
 	\label{eq:HER}
 \end{equation}
 \begin{equation}
 	\varepsilon _{q}(x,t)=\varepsilon _{q}(x)x^{-\alpha ^{\prime }(1-x)t}.
 	\label{eq:EER}
 \end{equation}

Additionally, the modified Gaussian ansatz (MG), which is as follows~\cite{Selyugin:2009ic,Sharma:2016cnf}:
\begin{equation}
	\mathcal{H}^{q}(x,t)=q_{v}(x)\exp \left[ \alpha \frac{(1-x)^{2}}{x^{m}}t%
	\right] ,
	\label{eq:HMG}
\end{equation}

\begin{equation}
	\varepsilon _{q}(x,t)=\varepsilon _{q}(x)\exp \left[ \alpha \frac{(1-x)^{2}}{x^{m}}t%
	\right] ,
	\label{eq:EMG}
\end{equation}

\begin{figure*}
	%\vspace*{-.658cm}
	\includegraphics[clip,width=0.45\textwidth]{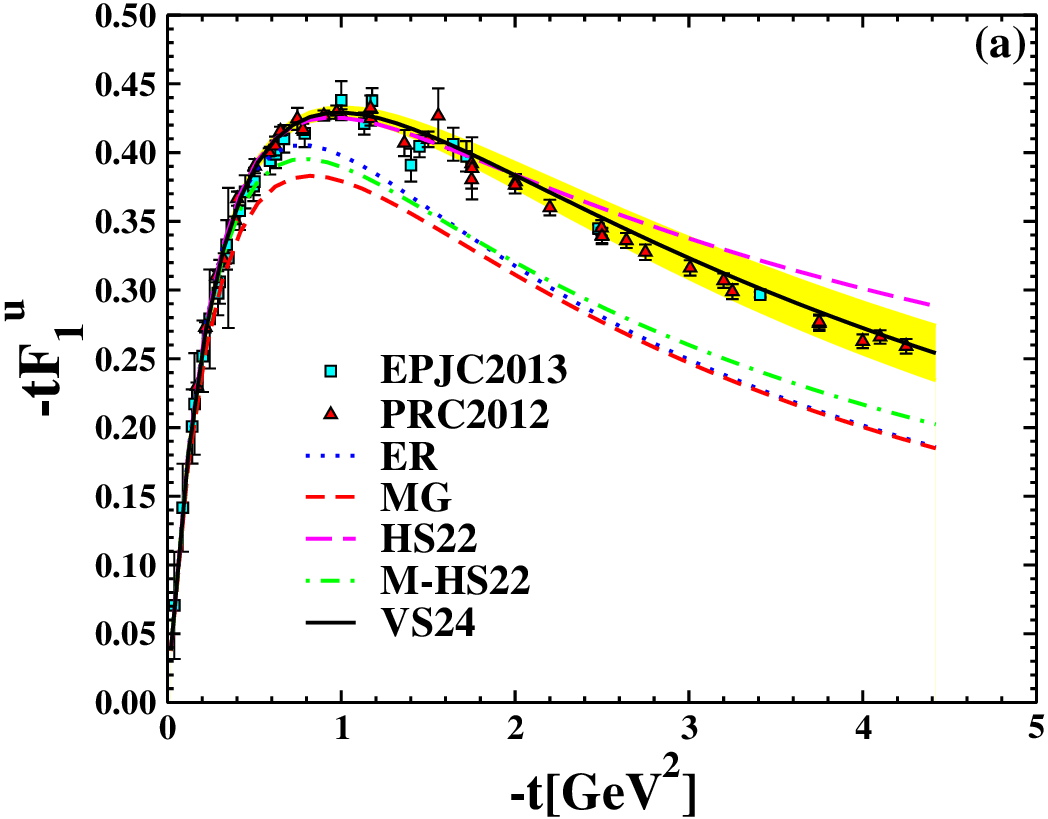}
	\hspace*{2mm}
	\includegraphics[clip,width=0.45\textwidth]{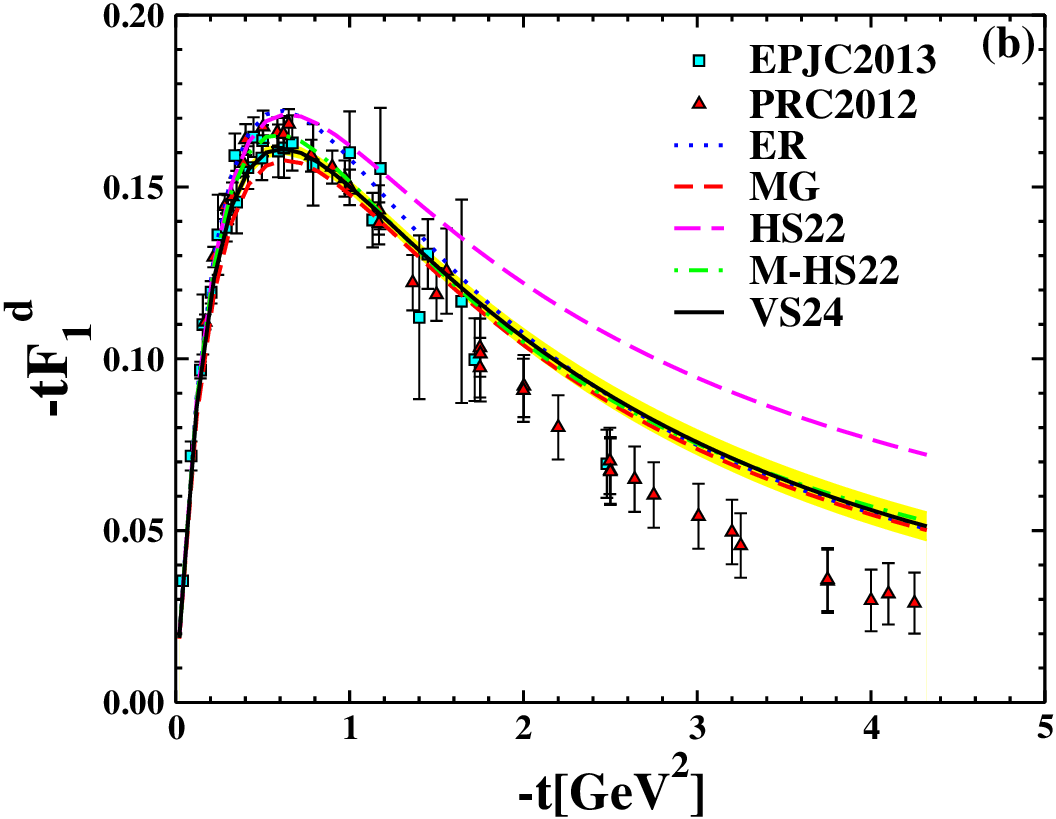}
	\vspace*{3mm}\\
	\includegraphics[clip,width=0.45\textwidth]{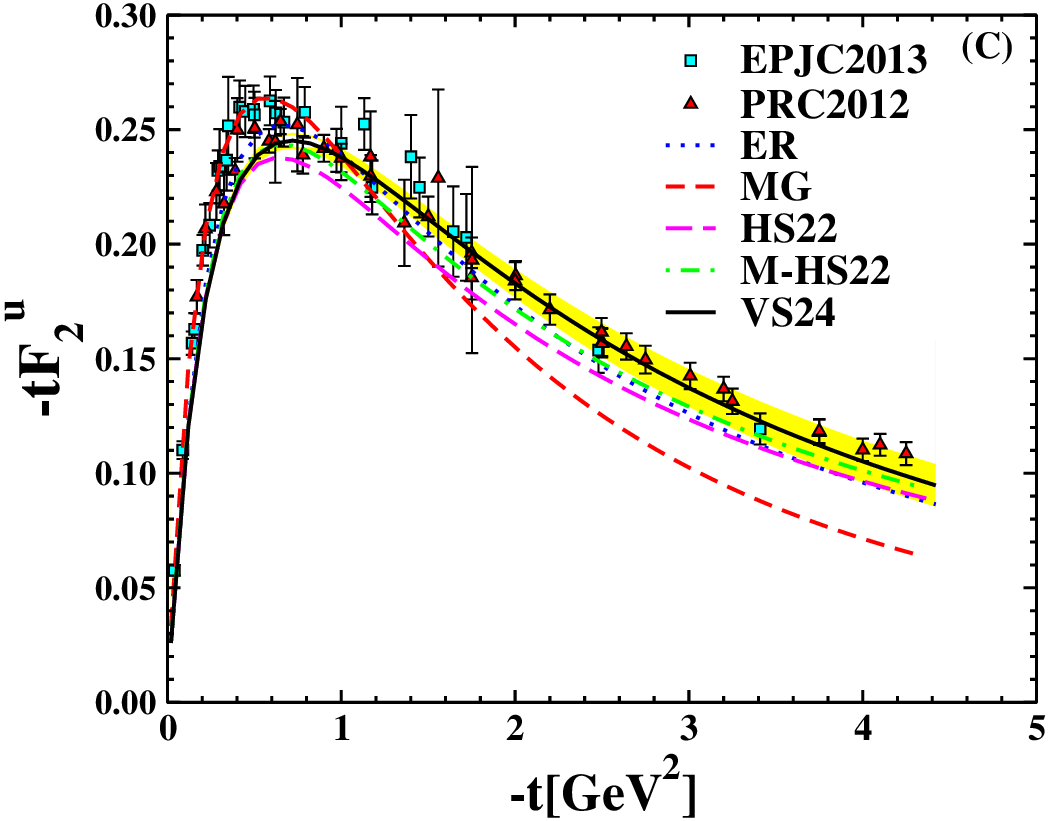}
	\hspace*{2mm}
	\includegraphics[clip,width=0.45\textwidth]{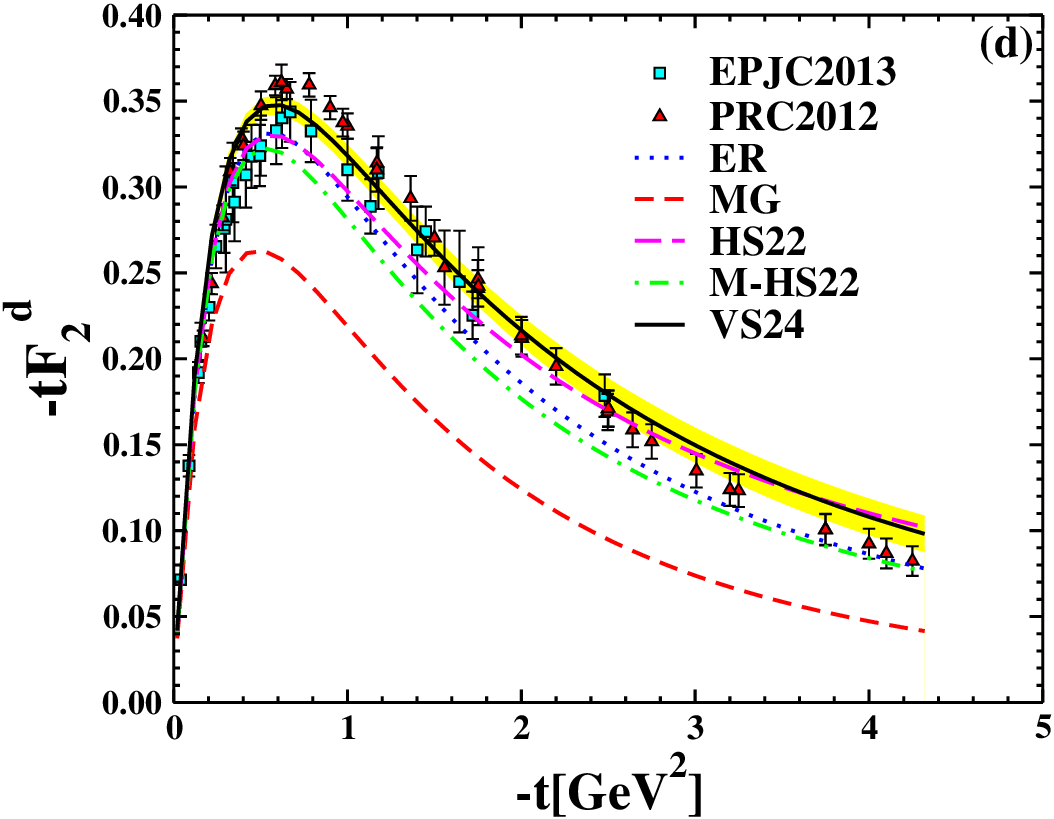}
	\vspace*{1.5mm}
	
	\caption{\footnotesize  The $F_1^{u,d}$ and $F_2^{u,d}$ multiplied by $t$ as a function of $-t$. Comoarsion of the ER ansatz~\cite{Guidal:2004nd}, the MG ansatz\cite{Selyugin:2009ic}, the HS22 ansatz~\cite{HajiHosseiniMojeni:2022tzn}, the M-HS22 ansatz ~\cite{Vaziri:2023xee} with the VS24 ansatz. The  KKA10 PDF~\cite{Khorramian:2009xz} are used in all of them. The extracted points are based on experimental data from ~\cite{Qattan:2012zf} (triangle up),~\cite{Cates:2011pz} (circle) and ~\cite{Diehl:2013xca} (square).}
	\label{fig:tfud1}
\end{figure*}

\begin{figure*}
	%\vspace*{-.658cm}
	\includegraphics[clip,width=0.45\textwidth]{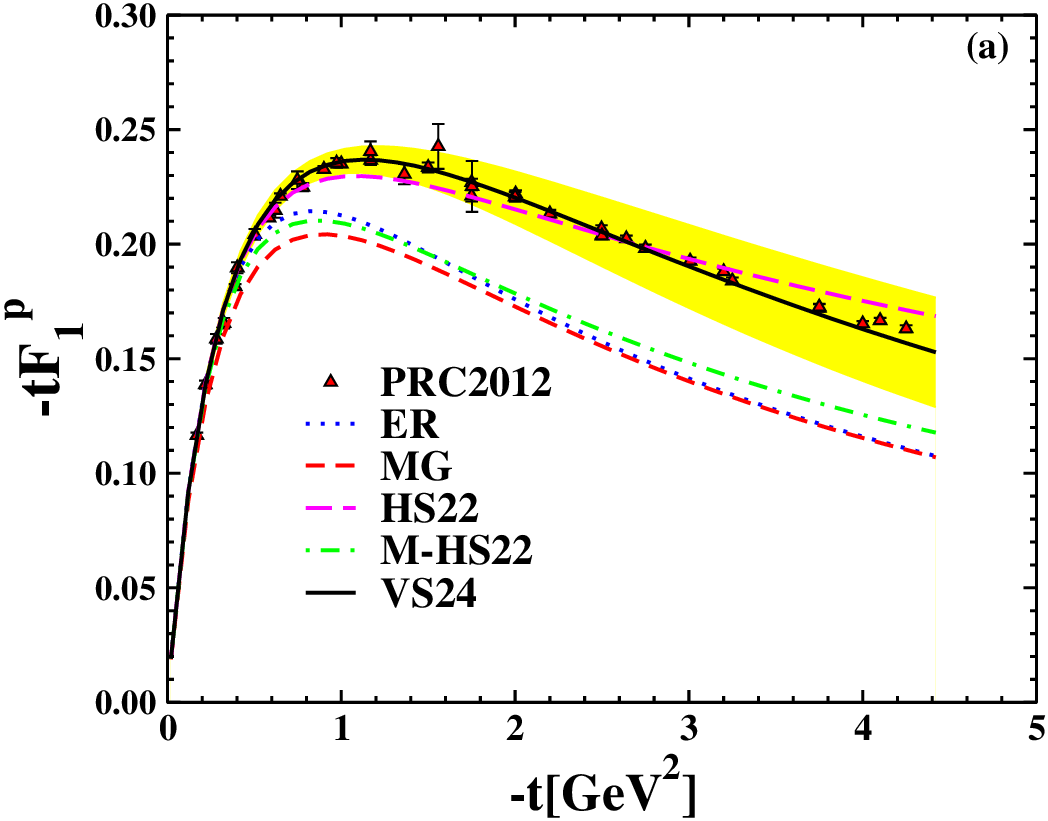}
	\hspace*{2mm}
	\includegraphics[clip,width=0.45\textwidth]{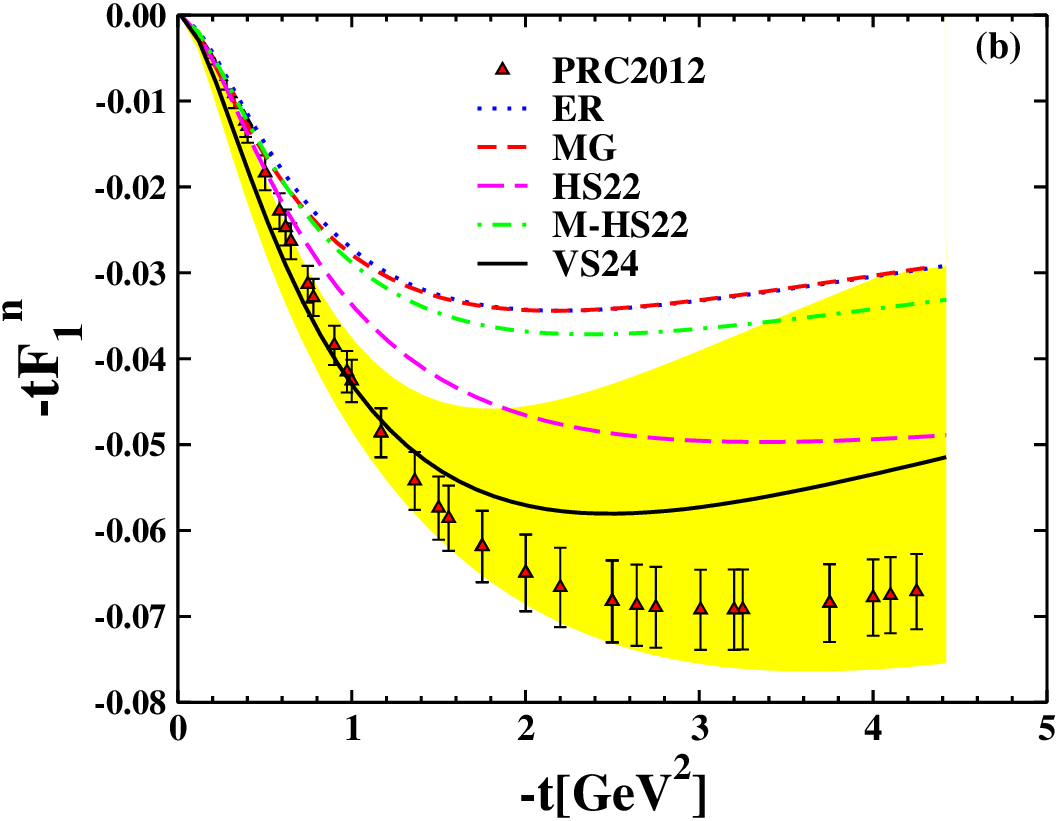}
	\vspace*{3mm}\\
	\includegraphics[clip,width=0.45\textwidth]{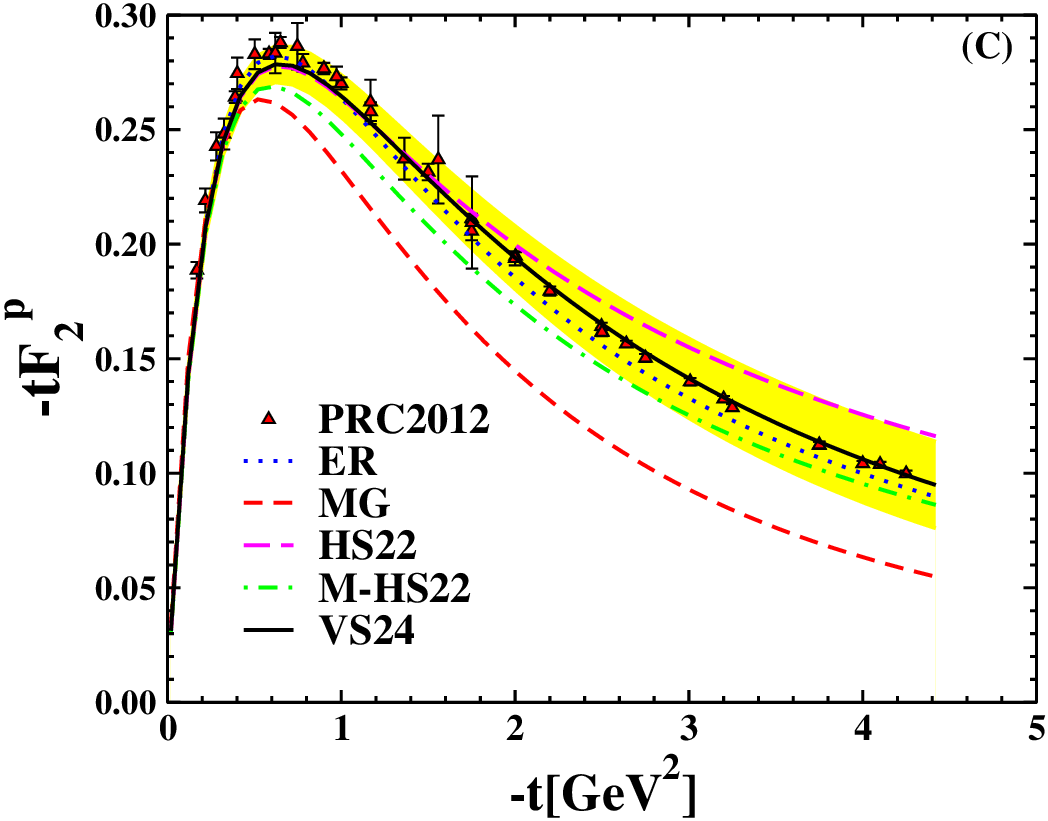}
	\hspace*{2mm}
	\includegraphics[clip,width=0.45\textwidth]{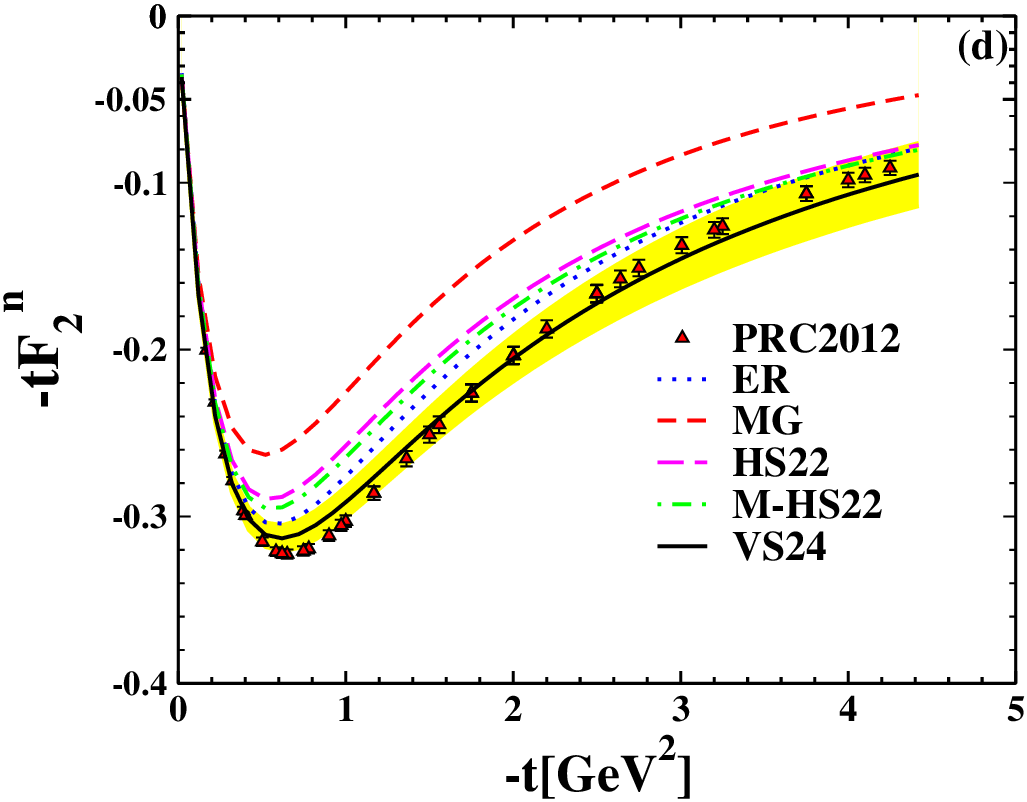}
	\vspace*{1.5mm}
	\caption{\footnotesize  The $F_1^{p,n}$ and $F_2^{p,n}$ multiplied by $t$ as a function of $-t$. Comoarsion of the ER ansatz~\cite{Guidal:2004nd}, the MG ansatz\cite{Selyugin:2009ic}, the HS22 ansatz~\cite{HajiHosseiniMojeni:2022tzn}, the M-HS22 ansatz ~\cite{Vaziri:2023xee} with the VS24 ansatz. The  KKA10 PDF~\cite{Khorramian:2009xz} are used in all of them. The extracted points are based on experimental data from ~\cite{Qattan:2012zf} (triangle up).}
	\label{fig:tfud}
\end{figure*}

\begin{figure*}
	%\vspace*{-.658cm}
	\includegraphics[clip,width=0.45\textwidth]{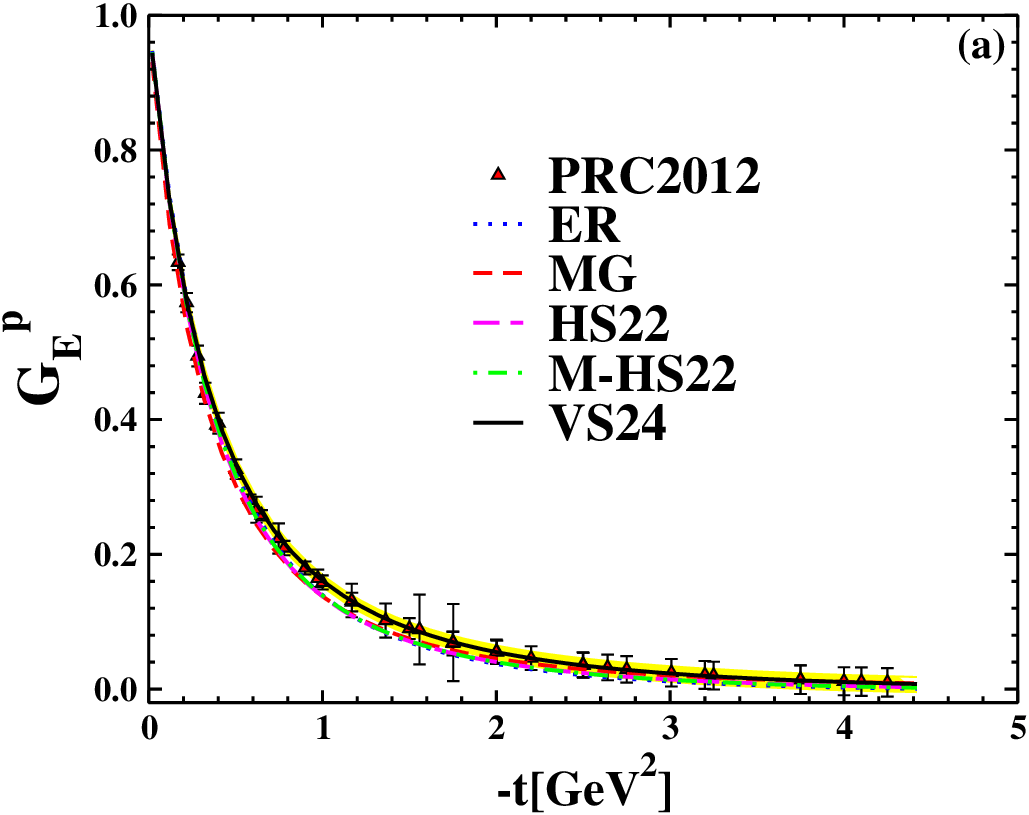}
	\hspace*{2mm}
	\includegraphics[clip,width=0.45\textwidth]{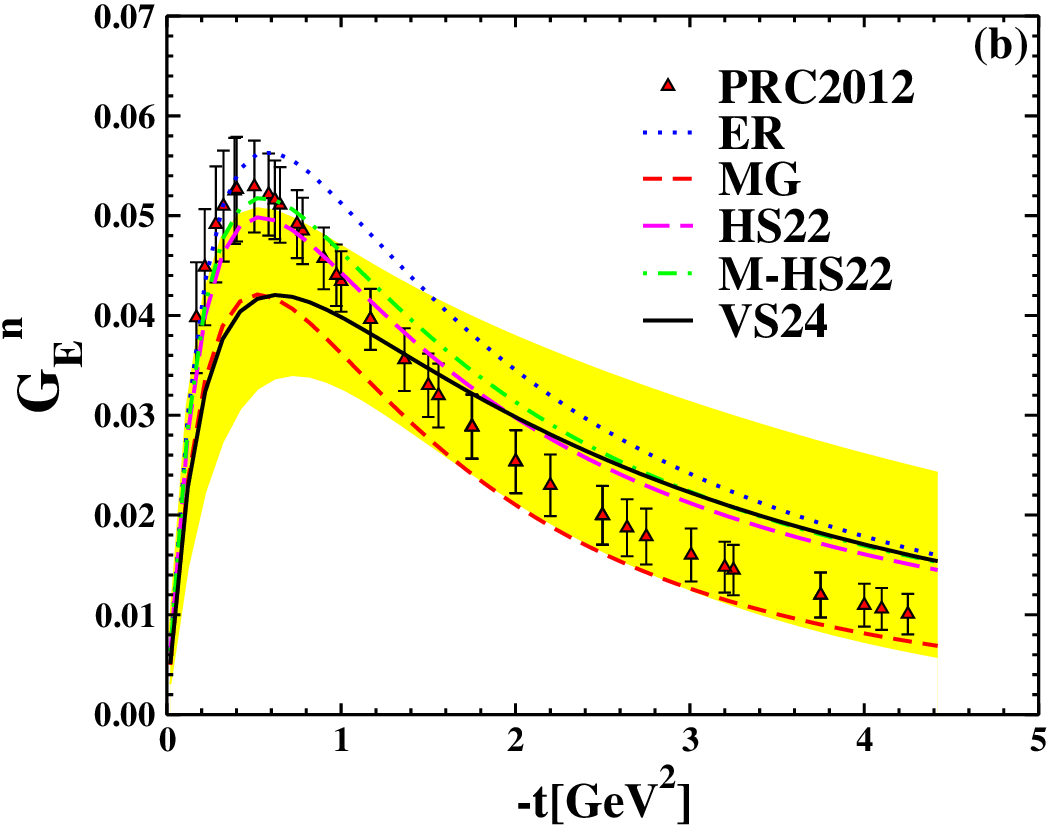}
	\vspace*{3mm}\\
	\includegraphics[clip,width=0.45\textwidth]{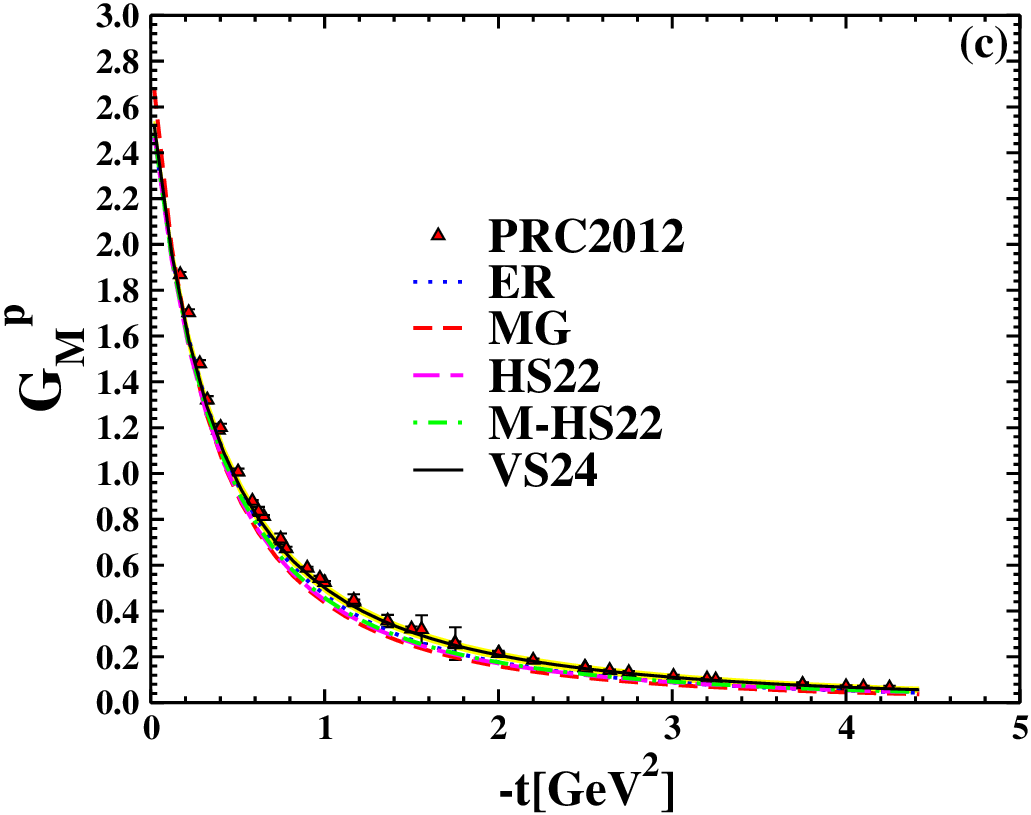}
	\hspace*{2mm}
	\includegraphics[clip,width=0.45\textwidth]{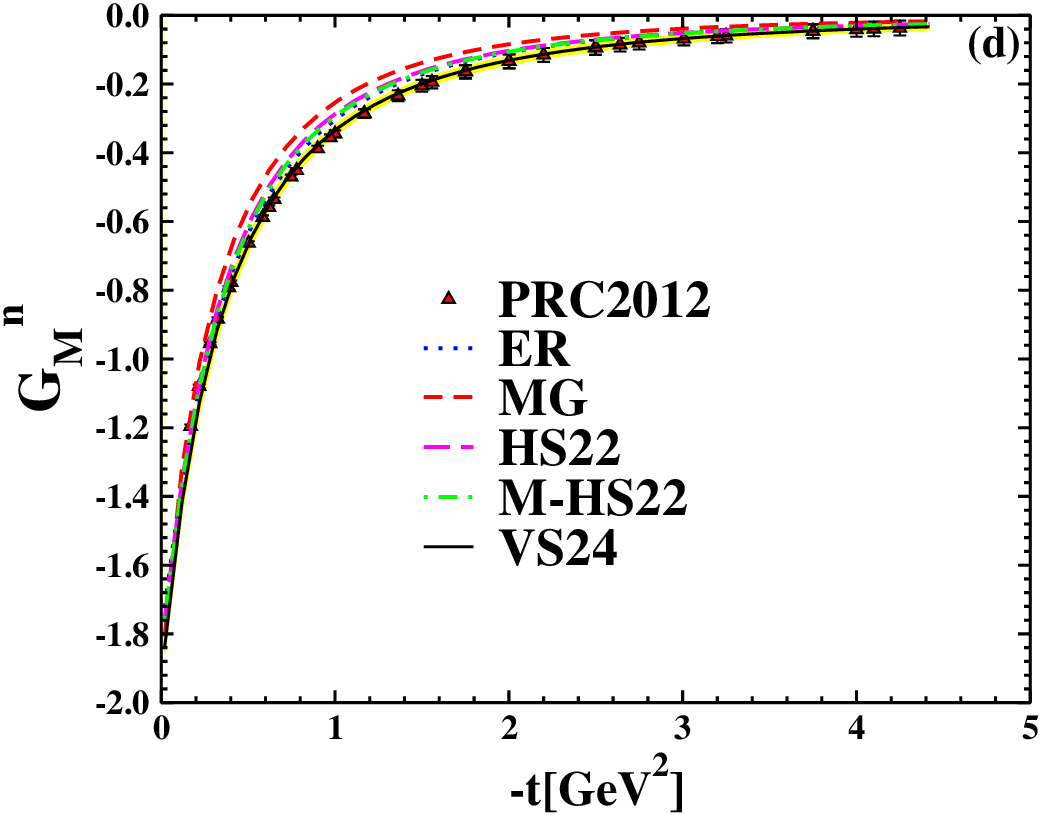}
	\vspace*{1.5mm}
	\caption{\footnotesize    The $G_E^{p,n}$ and $G_M^{p,n}$ as a function of $-t$. Comoarsion of the ER ansatz~\cite{Guidal:2004nd}, the MG ansatz\cite{Selyugin:2009ic}, the HS22 ansatz~\cite{HajiHosseiniMojeni:2022tzn}, the M-HS22 ansatz ~\cite{Vaziri:2023xee} with the VS24 ansatz. The  KKA10 PDF~\cite{Khorramian:2009xz} are used in all of them. The extracted points are based on experimental data from ~\cite{Qattan:2012zf} (triangle up).}
	\label{fig:GEMpn}
\end{figure*}

The calculated parameters that are free for the modified Gaussian (MG) and extended ER ansatzes are $m=0.45$, $\alpha^{\prime}=1.15$, and $\alpha=1.09$, respectively. These values are taken from~\cite{Selyugin:2009ic,Sharma:2016cnf} Furthermore, we introduce the HS22~\cite{HajiHosseiniMojeni:2022tzn} ansatz and the M-HS22~\cite{Vaziri:2023xee} ansatz, which provide an alternative approach to parametrize the GPDs. The HS22 and M-HS22 ansatz incorporate additional degrees of freedom and allow for a more comprehensive description of the nucleon structure. By considering this ansatz, we explored the impact of these additional parameters on the GPDs and their relationship to the nucleon's form factors.
\begin{equation}
	\mathcal{H}_{q}(x,t)=q_{v}\exp [-\alpha^{\prime \prime} t(1-x)\ln (x)+\beta x\ln (1-bt)],
	\label{eq:H}
\end{equation}

\begin{equation}
	\varepsilon _{q}(x,t)=\varepsilon _{q}(x)\exp [-\alpha^{\prime \prime} t(1-x)\ln (x)+\beta
	x\ln (1-bt)].
	\label{eq:E}
\end{equation}

We introduced changes to the parameters $\alpha^{\prime\prime}$, $\beta$, and $b$, which were taken to be 1.125, 0.185, and 1.82, respectively, for M-HS22 in Ref.~\cite{Vaziri:2023xee} and calculated the results at the $N^2LO$ approximation of PDF. But in this section, we introduce the VS24 ansatz and utilize this ansatz and their respective parameter values to analyze and interpret experimental data. By comparing the predictions of the VS24 model with experimental results, we can gain insights into the behavior of GPDs and their role in shaping the nucleon's structure as follows~\cite{HajiHosseiniMojeni:2022tzn,Vaziri:2023xee}:

\begin{equation}
	\mathcal{H}_{q}(x,t)=q_{v}\;\exp [-\alpha^{\prime \prime \prime} t(1-x)^\gamma\ln (x)+\beta x^{m^\prime}\ln (1-bt)],
	\label{eq:H}
\end{equation}
\begin{equation}
	\varepsilon _{q}(x,t)=\varepsilon _{q}(x)\;\exp [-\alpha^{\prime \prime \prime} t (1-x)^\gamma\ln (x)+\beta
	x^{m^\prime}\ln (1-bt)].
	\label{eq:E}
\end{equation}
In the modified VS24 ansatz, these parameter values have been chosen to reproduce the experimental data reported in Refs.~\cite{Diehl:2013xca,Cates:2011pz,Qattan:2012zf}. By incorporating the VS24 ansatz in this paper, we aim to investigate its impact on the results.

	 \begin{table}[h]
	 	\caption{{ Coefficient for Eqs.\ref{eq:H},\ref{eq:E}}.
	 		\label{tab:tab1}}
	 	\begin{tabular}{ll}
	 		\hline
	 		$\alpha ^{\prime \prime \prime }=1.3742\pm 7.16734\times 10^{-3},$ &$b=2$ \\
	 		$\beta =1.52378\pm 2.8392\times 10^{-2},$ & $m^{\prime }=0.65$  \\ 
	 		$\gamma=0.0570108\pm 1.49425\times 10^{-2}$   & \\ 
	 		$\eta{_u}=0.71207\pm 9.909\times 10^{-3}$   & \\ 
	 		$\eta{_d}=0.19248\pm 1.606\times 10^{-2}$   & \\ \hline
	 	\end{tabular}
	 \end{table}

	 Our study employs one distinct parton distribution for calculating the results, KKA10 PDF~\cite{Khorramian:2009xz}.\\
	 The KKA10 parton distribution functions in the NNNLO approximation are discussed for an input of $Q_0 = 4.0 GeV$~\cite{Khorramian:2009xz}: 
	 \begin{equation}
	 	xu_v=3.41356x^{0.80927}(1 - x)^{3.76847}(1 + 0.1399 x^{0.5} - 1.12 x),
	 \label{eq:KUV}
	 \end{equation}
 
 \begin{equation}
	 	xd_v=5.10129x^{0.79167}(1 - x)^{4.02637}(1 + 0.09 x^{0.5} + 1.11 x).
	 	\label{eq:KdV}
	 \end{equation}
 Using the VS24 ansatz in combination with KKA10 parton distribution functions (PDFs), we calculate the form factors of the $u$ and $d$ quarks based on the formalism described earlier. The results of the form factors that were obtained by different ansatzes are shown in Fig.~\ref{fig:tfud1}, and we analyzed the parameters of the VS24 ansatz. Furthermore, by considering four different ansatz combined with KKA10 PDF and performing the calculations, the Dirac and Pauli form factors of the proton and neutron, $F_1$ and $F_2$, are presented as functions of $-t$ in Fig.~\ref{fig:tfud}.
 
   It is observed that the combination of the VS24 ansatz and KKA10 PDF yields a better agreement with the experimental data that are in Ref.~\cite{Diehl:2013xca,Cates:2011pz,Qattan:2012zf}, particularly for the form nucleon's factors, compared to the other combinations mentioned above. In our calculations, the parameters of $b$ and $m ^{\prime}$ are fixed, and the other parameters in Tab.(\Ref{tab:tab1}), are determined by fitting data from ~\cite{Qattan:2012zf,Cates:2011pz,Diehl:2013xca}.

	\section{THE NUCLEON'S  MAGNETIC AND  ELECTRIC FORM FACTORS}\label{sec:sec4}
At present, the preferred electric and magnetic form factors for proton and neutron are $G_{E}(q^2)$ and $G_{M}(q^2)$. The Dirac $F_1^N$ and Pauli $F_2^N$ form factors are combined linearly ~\cite{Ernst:1960zza,Nikkhoo:2017won}:

\begin{equation}
	G_{E}^{N}(t)= -\tau F_{2}(t)+F_{1}(t)    ,\;\;\;G_{M}^{N}(t)=F_{2}(t)+F_{1}(t).
	\label{eq:GN}
\end{equation}
Where $\tau \equiv -t / 4 M_N^2$ and $t=Q^2$ is the four-momentum transfer of the virtual photon (the photon virtuality). According to Sach, the form factors $G_{E}$ and $G_{E}$ may have a more fundamental significance than $F_1$ and $F_2$ when it comes to a physical interpretation of these form factors in terms of spatial distributions of charge and magnetization inside the nucleon~\cite{Ernst:1960zza}. 

The electric and magnetic form factors, denoted by $G_{E}$ and $G_{M}$, respectively, and the experimental values of them are obtained through the Rosenbluth equation~\cite{Rosenbluth:1950yq}.

At $q^2=0$ , the calculated form factors in VS24 ansatz have been compared to the form factors obtained from scattering experiments in the Rosenbluth equation~\cite{Hohler:1976ax}, as shown in Tab.(\ref{tab:tabq}).

\begin{table}[h]
	\begin{center}
		\caption{{\footnotesize Comparing the computed form factors $G_{E}$ and $G_{M}$ between our results and calculations derived from experimental data ~\cite{Hohler:1976ax}. The findings are normalised at $q^2=0$.}
			\label{tab:tabq}}
		\vspace*{0.8mm}
		\begin{tabular}{ccccc}
			\hline\hline\
			\\	Ansatz & {\hspace{1mm}}$G_{E}^{P}(0)$ & {\hspace{1mm}}$G_{M}^{P}(0)$& {\hspace{1mm}}$G_{E}^{n}(0)$& {\hspace{1mm}}$ G_{M}^{n}(0)$   \\   \hline
			\\Exprimental data   & {\hspace{4mm}}$1$ & {\hspace{4mm}}$+2.79$& {\hspace{4mm}}$0$& {\hspace{4mm}}$ -1.91$  \\
			VS24   & {\hspace{4mm}}$1$ & {\hspace{4mm}}$+2.658$& {\hspace{4mm}}$1.1\times 10^{-13}$& {\hspace{4mm}}$ -1.944$  \\   
			
			\hline						
		\end{tabular}
	\end{center}
\end{table}

Since a neutron has no charge, the sign of the electric neutron form factor cannot be determined in this manner. However, the electromagnetic form factors of the neutron approximately follow the dipole form~\cite{Miller:1990iz}.
 It is evident that the errors on $F_1$ and $F_2$ are typically more correlated and larger compared to the errors on $G_E$ and $G_M$.
 
   The cross section's preferred is expressed in terms of $G_E$ and $G_M$~\cite{Hand:1963zz}.

This finding is consistent with a simple, nonrelativistic interpretation where the charge and magnetization of the nucleon are carried by its constituent quarks. Specifically, the up and down quarks exhibit similar spatial distributions, leading to similar contributions to all form factors, except for $G^n_E$. In the case of $G^n_E$, there is a nearly complete cancellation between the contributions from the up and down quark charge distributions. The precise measurements of $G^n_E$ have played a crucial role in establishing. There are differences between the down- and up-quark distributions~\cite{Drell:1969km}:

\begin{eqnarray}
	G^p_{E,M}(Q^2)&=&\frac{2}{3}G^u_{E,M}(Q^2)-\frac{1}{3}G^d_{E,M}(Q^2),\nonumber\\
	G^n_{E,M}(Q^2)&=&\frac{2}{3}G^d_{E,M}(Q^2)-\frac{1}{3}G^u_{E,M}(Q^2).
\end{eqnarray}

	By studing the up-and-down quarks' contributions to the nucleon's form factors, we can learn  more about the  fundamental structure and dynamics of the nucleon~\cite{Beck:2001yx}. In this convention, the form factors $G^u_{E,M}$ represent the contributions of up quarks in the proton and down quarks in the neutron to the electric and magnetic form factors, respectively~\cite{Perdrisat:2006hj}. The nucleon's electric and magnetic form factors are shown in Fig.~\ref{fig:GEMpn}, derived from various ansatz models and KKA10 parton distribution functions~\cite{Khorramian:2009xz}. These form factors are plotted as functions of $-t$, the squared momentum transfer, and then the calculations of them are analyzed.

	\section{The ELECTRIC RADII OF THE NUCLEONS BASED ON VS24 ANSATZ}\label{sec:sec5}
	
At zero momentum transfer, the form factors' slope determines the particle radii, and the squares of the Dirac radius $<r_{D}^2>$ and charge radius $r_{E,p}^2$ are ascertained as follows~\cite{Selyugin:2019dav}: 
	 	 
\begin{eqnarray}
	<r_{D}^2>&=&-6 \frac{dF^{p,n}_{1}(t)}{dt} \mid_{t=0},\nonumber\\	 
	<r_{E}^2>&=&-6 \frac{dF^{p,n}_{1}(t)}{dt} \mid_{t=0}+\; \frac{3}{2} \frac{{\kappa_{{n},p}}}{m_{n,p}^2}.
	 \label{eq:r1,p}
\end{eqnarray}

Furthermore, we calculated the Dirac mean squared radii of the nucleon using the extended Regge ansatz (ER)~\cite{SattaryNikkhoo:2018gzm}.
\begin{equation}
	<r_{D,p}^{2}>=-6\alpha ^{\prime
	}\int_{0}^{1}dx[e_{u}u_{v}(x)+e_{d}d_{v}(x)] \ln (x) (1-x),\label{eq:rpREEGE}
\end{equation}

\begin{equation}
	<r_{D,n}^{2}>=-6\alpha ^{\prime
	}\int_{0}^{1}dx[e_{u}d_{v}(x)+e_{d}u_{v}(x)] \ln (x) (1-x).\label{eq:rnREEGE1}
\end{equation}

Moreover, the MG model is used to compute both the neutron and proton electric radii:
\begin{equation}
	<r_{D,p}^{2}>=-6\alpha \int_{0}^{1}dx[e_{u}u_{v}(x)+e_{d}d_{v}(x)]\;\frac{%
		(1-x)^{2}}{x^{m}},\label{eq:rPMG}
\end{equation}

\begin{equation}
	<r_{D,n}^{2}>=-6\alpha \int_{0}^{1}dx[e_{u}d_{v}(x)+e_{d}u_{v}(x)]\;\frac{%
		(1-x)^{2}}{x^{m}}.\label{eq:rPMG1}
\end{equation}

And Finally for VS24 ansatz, we have got:

\begin{eqnarray}
	<r_{D,p}^{2}>=-6\alpha^{\prime \prime\prime}\int_{0}^{1}dx    
	[e_{u}u_{v}(x)+e_{d}d_{v}(x)]   (1-x)^\gamma\nonumber\\ 
	\ln (x)+\beta x^{m^\prime}\ln(1-bt),\;\;\;\;\;\;\;\;\;\;\;\;\;\;\;\;\;\;\;\;\;\;\;\;\;\;\;
	\label{eq:rPMG2}
\end{eqnarray}

\begin{eqnarray}
	<r_{D,n}^{2}>=-6\alpha^{\prime \prime\prime}\int_{0}^{1}dx    
	[e_{u}d_{v}(x)+e_{d}u_{v}(x)]   (1-x)^\gamma\nonumber\\ 
	\ln (x)+\beta x^{m^\prime}\ln(1-bt).\;\;\;\;\;\;\;\;\;\;\;\;\;\;\;\;\;\;\;\;\;\;\;\;\;\;\;
	\label{eq:rPMG3}
\end{eqnarray}

In Tab.(\ref{tab:tabR}), we present the computed electric radii of nucleons using various generalized parton distributions (GPDs) and compare them with experimental data obtained from \cite{ParticleDataGroup:2010dbb}.
\captionsetup{belowskip=0pt,aboveskip=0pt}

\begin{table}[h]
	\begin{center}
		\caption{{\footnotesize The nucleon's electric radii were determined by employing the KKA10 parton distribution functions (PDFs)~\cite{Khorramian:2009xz} and different ansatzes, namely the extended (ER) \cite{Guidal:2004nd}, (MG) \cite{Selyugin:2009ic},HS22~\cite{{HajiHosseiniMojeni:2022tzn}}, MHS22~\cite{{Vaziri:2023xee}} and VS24 models. The data utilized in this research are sourced from the Ref. \cite{ParticleDataGroup:2010dbb}.}
			\label{tab:tabR}}
		\vspace*{1.5mm}
		\begin{tabular}{ccc}
			\hline\hline\
			\\	Ansatz & {\hspace{10mm}}$r_{E,p}$ & {\hspace{10mm}}$r^2_{E,n} $   \\   \hline
			\\Exprimental data   &{\hspace{10mm}}   0.877 $fm$ &{\hspace{5mm}} -0.1161 $fm^2$ \\
			VS24   &{\hspace{10mm}}   0.834 $fm$ &{\hspace{5mm}} -0.1558 $fm^2$ \\   
			ER   &{\hspace{10mm}}   0.811 $fm$ &{\hspace{5mm}} -0.0849 $fm^2$ \\ 
			MG   &{\hspace{10mm}} 0.952 $fm$ &{\hspace{5mm}} -0.09386 $fm^2$ \\ 
			HS22   &{\hspace{10mm}} 0.833 $fm$ &{\hspace{5mm}} -0.07229 $fm^2$ \\
			M-HS22&{\hspace{10mm}}   0.827 $fm$ &{\hspace{5mm}} -0.0723 $fm^2$ \\
			\hline\hline						
		\end{tabular}
	\end{center}
\end{table}

	\section{ RESULTS AND CONCLUSION
	}\label{sec:conclusion}
	The study of hadrons' internal structure has been studied in the past decade using the GPDs at $LO$, $NLO$, and $N^2LO$ approximations. The results for the nucleon's Pauli and Dirac form factors are shown in this section, as well as the electric and magnetic form factors of both and the flavor components of the nucleon. These results were obtained by employing the ansatz in GPDs (VS24 model) and utilizing the KKA10 parton distributions at $N^3LO$ approximation. We have plotted the form factors, analyzed the parameters of the ansatz in graphs as a function of $-t$, then calculated the radii of the proton and neutron, which are then presented in a table. We started the task by introducing the form factor equations in Sec. (\ref{sec:sec2}) and used Eqs. (\ref{eq:F2}) and (\ref{eq:F3}) to calculate them. Considering that the GPDs can be obtained from Eqs. (\ref{eq:F1}-- \ref{eq:F4}) , which consist of two parts, the first being the ansatz and dependent on three quantities, $x$, $t$, and $\xi$; the second part of the equation also includes a PDF that is dependent on $x$. Based on the results of the form factor calculations in this paper, since the effect of the ansatz is greater than the PDF, we combined four ansatzes: the ER ansatz~\cite{Guidal:2004nd}, the MG ansatz\cite{Selyugin:2009ic}, the HS22 ansatz~\cite{HajiHosseiniMojeni:2022tzn}, and the M-HS22 ansatz~\cite{Vaziri:2023xee} with one PDF at $N^3LO$ approximation and compared the results with the combination of the PDF and VS24 ansatz. We proposed VS24 model in Eqs. (\ref{eq:H}) and (\ref{eq:E}) that parameters of $b$ and $m ^{\prime}$ in this ansatz are fixed, and the other parameters $\alpha ^{\prime \prime \prime}$, $\beta$, and $\gamma$ are determined by fitting data from ~\cite{Qattan:2012zf,Cates:2011pz,Diehl:2013xca}. Then we introduced in Eqs. (\ref{eq:KUV}) and (\ref{eq:KdV}) KKA10 parton distribution functions that are at the $N^3LO$ approximation ~\cite{Khorramian:2009xz}. The quality of the fits is shown in Tab. (\ref{tab:tab1}). Combination of the PDF and ansatz that mentioned, we calculated the form factors of $u$ and $d$ quarks, as well as for protons and neutrons. These calculations were then analyzed, and their graphs were plotted as a function of $-t$, respectively, in Figs. \ref{fig:tfud1} and \ref{fig:tfud}. In these figures, the results of calculations for other ansatzes in combination with KKA10 PDF are plotted at $N^3LO$ approximation. It is evident that the combination of VS24-KKA10 has a closer match with the form factors extracted from the data of electron scattering experiments from protons~\cite{Qattan:2012zf,Cates:2011pz,Diehl:2013xca}. As Figs. (\ref{fig:tfud1}--\ref{fig:GEMpn}) are shown, our calculations indicate that for the four ansatzes that were mentioned, the changes in the form factors as a function of $t$ in the $N^3LO$ approximation are closely in agreement with the experimental data in the small-x region only. But the VS24 ansatz for all of the x ranges for both the proton and neutron is in agreement with the experimental data. Considering that the study of the electric and magnetic form factors of the nucleons provides us with more information about the internal structure of hadrons compared to the Pauli and Dirac form factors, we addressed this topic in Sec. (\ref{sec:sec4}). The $G_E^{p,n}$ and $G_M^{p,n}$ as a function of $-t$ are shown in Fig.\ref{fig:GEMpn}, which compares the four ansatz with the VS24 ansatz. This result was obtained by using Eqs. (\ref{eq:GN}). We have analyzed the parameters of the VS24 ansatz and plotted them for the upper and lower bounds in all figures of the paper. Additionally, we have calculated these form factors in $q^2=0$ using the desired ansatz and compared them with the form factors calculated from the exprimental data of Ref.~\cite{Hohler:1976ax}. In Tab. (\ref{tab:tabq}), the computation results are displayed.
		
One of the important applications of form factors is to calculate the electric radii of protons and neutrons using them; therefore, in Sec. (\ref{sec:sec5}), we discuss the nucleon radii. The nucleon's Dirac mean squared radii are calculated with the extended Regge ansatz (ER)~\cite{Guidal:2004nd}, the MG ansatz\cite{Selyugin:2009ic}, and the VS24 model are calculated by Eqs. (\ref{eq:rpREEGE}-- \ref{eq:rPMG3}). The results of these calculations for five types of ansatz are shown in Tab. (\ref{tab:tabR}). The chosen parametrizations for the VS24 ansatz provide a suitable match with the data used in Ref. \cite{ParticleDataGroup:2010dbb}.

	\section{Data Availability Statement:No Data associated
	}\label{sec:conclusion1}

	%%%%%%%%%%%%%%%%%%%%%%%%%%%%%%%%%%%%%%%%%%%%%%%%%%%%%%%%%

	%\clearpage

	%
	%%%%%%%%%%%%%%%%%%%%%%%%%%%%%%%%
	
\end{document}